# The metallicity dependence of the CO → H$_2$ conversion factor in z≥1 star forming galaxies[1]

R.Genzel[1,2], L.J.Tacconi[1], F.Combes[3], A.Bolatto[4], R.Neri[5], A.Sternberg[6], M.C.Cooper[7], N.Bouché[8], F.Bournaud[9], A.Burkert[10,11], J.Comerford[12], P.Cox[5], M.Davis[13], N.M. Förster Schreiber[1], S.Garcia-Burillo[14], J.Gracia-Carpio[1], D.Lutz[1], T.Naab[15], S.Newman[13], A.Saintonge[1,15], K.Shapiro[16], A.Shapley[17] & B.Weiner[18]

[1] *Max-Planck-Institut für extraterrestrische Physik (MPE), Giessenbachstr.1, 85748 Garching, Germany*

   *( genzel@mpe.mpg.de,  linda@mpe.mpg.de )*

[2] *Dept. of Physics, Le Conte Hall, University of California, CA 94720 Berkeley, USA*

[3] *Observatoire de Paris, LERMA, CNRS, 61 Av. de l'Observatoire, F-75014 Paris, France*

[4] *Dept. of Astronomy, University of Maryland, College Park, MD 20742-2421, USA*

[5] *IRAM, 300 Rue de la Piscine, 38406 St.Martin d'Heres, Grenoble, France*

[6] *Sackler School of Physics and Astronomy, Tel Aviv University, Tel Aviv 69978, Israel*

[7] *Hubble Fellow, Department of Physics & Astronomy, Frederick Reines Hall, University of  California, Irvine, CA 92697-4575, USA*

[8] *Dept. of Physics, University of California, Santa Barbara, Broida Hall, Santa Barbara CA 93106, USA*

[9] *Service d'Astrophysique, DAPNIA, CEA/Saclay, F-91191 Gif-sur-Yvette Cedex, France*

[10] *Universitätssternwarte der Ludwig-Maximiliansuniversität , Scheinerstr. 1, D-81679 München, Germany*

[11] *MPG-Fellow at MPE*

[12] *Department of Astronomy & McDonald Observatory, 1 University Station, C1402 Austin, Texas 78712-0259, USA*

[13] *Department of Astronomy, Campbell Hall, University of California, Berkeley, CA  94720, USA*

[14] *Observatorio Astronómico Nacional-OAN, Apartado 1143, 28800 Alcalá de Henares- Madrid, Spain*

---

[1] Based on observations with the Plateau de Bure millimetre interferometer, operated by the Institute for Radio Astronomy in the Millimetre Range (IRAM), which is funded by a partnership of INSU/CNRS (France), MPG (Germany) and IGN (Spain).

[15] *Max-Planck Institut für Astrophysik (MPA), Karl Schwarzschildstrasse 1, D-85748 Garching, Germany*

[16] *Aerospace Research Laboratories, Northrop Grumman Aerospace Systems, Redondo Beach, CA 90278, USA*

[17] *Department of Physics & Astronomy, University of California, Los Angeles, CA 90095-1547, USA*

[18] *Steward Observatory, 933 N. Cherry Ave., University of Arizona, Tucson AZ 85721-0065, USA*

## Abstract

We use the first systematic samples of CO millimeter emission in $z \geq 1$ 'main-sequence' star forming galaxies (SFGs) to study the metallicity dependence of the conversion factor $\alpha_{CO}$, from CO line luminosity to molecular gas mass. The molecular gas depletion rate inferred from the ratio of the star formation rate (SFR) to CO luminosity, is ~1 Gyr$^{-1}$ for near-solar metallicity galaxies with stellar masses above $M_S \sim 10^{11}$ $M_\odot$. In this regime the depletion rate does not vary more than a factor of two to three as a function of molecular gas surface density, or redshift between $z \sim 0$ and 2. Below $M_S$ the depletion rate increases rapidly with decreasing metallicity. We argue that this trend is not caused by starburst events, by changes in the physical parameters of the molecular clouds, or by the impact of the fundamental metallicity-SFR-stellar mass relation. A more probable explanation is that the conversion factor is metallicity dependent and that star formation can occur in 'CO-dark' gas. The trend is also expected theoretically from the effect of enhanced photodissociation of CO by ultraviolet radiation at low metallicity. From the available $z \sim 0$ and $z \sim 1-3$ samples we constrain the slope of the $\log(\alpha_{CO})$ –log (metallicity) relation to range between -1 and -2, fairly insensitive to the assumed slope of the gas-star formation rate relation. Because of the lower metallicities near the peak of the galaxy formation activity at $z \sim 1-2$ compared to $z \sim 0$, we suggest that molecular gas masses estimated from CO luminosities have to be substantially corrected upward for galaxies below $M_S$.

*Subject Headings: galaxies: evolution – galaxies: high redshift – galaxies: ISM – stars: formation – ISM: molecules*

# 1. Introduction

The evolution of galactic star formation as a function of cosmic time is driven by the complex interplay of interstellar gas components and their chemical evolution, stars and their radiation and feedback, star formation processes and galactic/intergalactic environments. In the Milky Way and nearby galaxies most or all star formation occurs in dense, cool giant molecular clouds (GMCs: Solomon et al. 1987, Young & Scoville 1991, Blitz 1993, McKee & Ostriker 2007, Bigiel et al. 2008, Leroy et al. 2008, Bolatto et al. 2008, Schruba et al. 2011). The most commonly used tracer of the $H_2$ molecule, the elusive building block of GMCs, is line emission from low-lying rotational transitions of $^{12}CO$. This is perhaps surprising since these transitions are optically thick ($\tau_{CO\ 1-0} \geq 10$) in typical GMCs. The information on gas mass is mainly contained in the width of the line if the gas motions in GMCs are virialized and if the line emission in a given object is the superposition of a number of such virialized clouds (Dickman et al. 1986, Solomon et al. 1987). As a result the relationship between velocity integrated line flux $F_{CO\ J}$ (Jy km/s) in the $J \rightarrow J-1$ transition and the total molecular gas mass (including 36% helium) on large scales is traditionally given by the empirical relation (Dickman et al. 1986, Appendix A in Tacconi et al. 2008, Obreschkow & Rawlings 2009)

$$M_{H_2+He} = \alpha_{CO\ 1-0} \times L'_{CO\ 1-0}$$

$$= 3.66x10^8 \times \alpha_{CO\ 1-0} \times \left(\frac{F_{CO\ J}}{Jy\ km/s}\right) \times (R_{1J}) \times (1+z)^{-3} \times \left(\frac{\lambda_{obs\ J}}{mm}\right)^2 \times \left(\frac{D_L}{Gpc}\right)^2 \quad (M_\odot) \quad (1),$$

where

$$\alpha_{CO\ 1-0} = h\left(\frac{(<n(H_2)>)^{1/2}}{T_{R\ 1}}\right) \times g(Z) \quad (2).$$

Here $L'_{CO\ 1-0}$ (K km/s pc$^2$) is the integrated line luminosity of the 1-0 CO line, $R_{1J}$ is the ratio of the product of the Rayleigh-Jeans brightness temperature $T_R$ and the beam filling factor in the $1 \rightarrow 0$ transition to the same product in the $J \rightarrow J\text{-}1$ transition (at the same angular resolution), $\lambda_{obs\ J}$ is the observed wavelength of the $J \rightarrow J\text{-}1$ transition and $D_L$ is the luminosity distance of the source[2]. The functions h and g encapsulate the dependence of the function $\alpha_{CO\ 1-0}$ (M$_\odot$/ (K km/s pc$^2$)), commonly called 'conversion factor', on the physical conditions of the interstellar medium (ISM), and on the metallicity Z. The conversion factor depends on the ratio of the square-root of the mean hydrogen density $<n(H_2)>$ and the Rayleigh-Jeans brightness temperature[2]. It also depends on the spatial distribution and mass fraction of the molecular gas (relative to stars) in the cloud, and potentially on other parameters, such as the magnitude of turbulence in the GMCs etc. (Downes et al. 1993, Downes & Solomon 1998, Obreshkov et al. 2009, Tacconi et al. 2008, Pelupessy & Papadopoulos 2009, Shetty et al. 2011a,b, Shetty et al. 2011a,b, Narayanan et al. 2011b, Glover & Mac Low 2011. Feldman, Gnedin & Kravtsov 2011). Since the penetration depth of external ultraviolet (UV) radiation destroying molecules

---

[2] For a transition J→J-1 at frequency $\nu_J$, excitation temperature $T_{ex}$ and optical depth $\tau_J$ the Rayleigh-Jeans brightness temperature is given by $T_{R\ J} = h\nu_J/k\ (exp(h\nu_J/(kT_{ex}))-1)^{-1}(1-exp(-\tau_J))$

depends on the extinction through the cloud, and thus on its metallicity, the optical depth and the effective conversion factor in CO transitions are predicted to depend on metallicity, especially in low metallicity, diffuse gas (van Dishoeck & Black 1986, 1988, Maloney & Black 1988, Sternberg & Dalgarno 1995, Hollenbach & Tielens 1999, Wolfire et al. 1990, 2010, Pelupessy & Papadopoulos 2009, Shetty et al. 2011a, Glover & Mac Low 2011).

In the galaxy-integrated ISMs of the Milky Way and nearby SFGs with near solar metallicity, as well as in dense star forming clumps of lower mass, lower metallicity galaxies, the empirical CO 1-0 conversion factors determined with dynamical, dust and γ-ray calibrations are consistent with a single value of $\alpha_{CO\ 1-0} = \alpha_G = 4.36 \pm 0.9$ $M_\odot$/(K km/s pc$^2$) (Strong & Maddox 1996, Dame, Hartmann & Thaddeus 2001, Grenier et al. 2005, Bolatto et al. 2008, Leroy et al. 2011, Abdo et al. 2010). In these environments GMCs appear to have similar physical properties and the functions h and g do not vary much (Bigiel et al. 2008, 2011, Leroy et al. 2008). In star forming regions or starburst galaxies the higher average densities drive $\alpha_{CO\ 1-0}$ upward. However, when combined with the simultaneous decrease in $\alpha_{CO\ 1-0}$ due to the higher gas temperatures caused by stellar heating in the same regions, the overall conversion factor $\alpha_{CO\ 1-0}$ fortuitously does not change more than a factor of two even in these cases. More significant deviations of $\alpha_{CO\ 1-0}$ from the Galactic value occur in extreme merger-driven starbursts, where the Galactic conversion factor appears to overestimate the true gas masses ($\alpha_{CO\ 1-0} \leq 1$, Solomon et al. 1997, Scoville et al. 1997, Downes & Solomon 1998, Tacconi et al. 2008). In the outer part of the Milky Way and other z~0 disk galaxies, as well as in low metallicity dwarf galaxies the Galactic conversion factor appears to underestimate the true molecular hydrogen content (e.g. Strong et al. 2004, Abdo et al. 2010, Schruba et al. 2011). This suggests that in z~0 SFGs $\alpha_{CO\ 1-0}$ scales with gas-phase oxygen abundance $Z_O \equiv 12+\log\{O/H\}$ as $Z_O^{-0.7....-2}$ (Rubio, Lequeux & Boulanger 1993, Wilson 1995, Arimoto et al. 1996, Israel 1997, 2000, Boselli et al. 2002, Leroy et al. 2011, Bolatto et al. 2011). However, a much shallower dependence of $\alpha_{CO}$ on $Z_O$ is inferred for dense extragalactic star forming clouds, presumably because their higher column densities are sufficient to maintain a large optical depth in the UV continuum and the CO 1-0 line, despite the lower oxygen abundances (e.g. Bolatto et al. 2008).

In this paper we present a pilot study of the dependence of $\alpha_{CO\ 1-0}$ on metallicity at high redshift, based on the first systematic measurements of CO emission in several samples of massive z>1 SFGs. We find that the application of a Galactic conversion factor underestimates molecular masses in some of these systems by factors between 2 and 10. The outliers are low metallicity galaxies. We propose a first order empirical relation to correct the CO→ $H_2$ conversion factor at z≥1 for this metallicity effect.

Throughout the paper we use a standard WMAP ΛCDM cosmology (Komatsu et al. 2011) and a Chabrier (2003) initial stellar mass function. To convert the 3-2 and 2-1 line fluxes to 1-0 fluxes we take $R_{13}$=2 and $R_{12}$=1.16 from recent empirical calibrations (Weiss et al. 2007, Dannerbauer et al. 2009, Ivison et al. 2011, Riechers et al. 2010), with the exception of the more compact eyelash, where Danielson et al. (2010) find $R_{13}$=1.5 (see Genzel et al. 2010 for more details). The total molecular gas masses in this paper computed with $\alpha_G$=4.36 (which includes a 1.36 upward correction for helium) are the same as those in Genzel et al. (2010), who use $\alpha_G$=3.2 (but do not include the helium correction in the conversion factor).

## 2. Properties of the observed galaxies

### 2.1 Galaxy sample

In this paper we discuss recent galaxy integrated measurements of the $^{12}$CO 3-2 and 2-1 lines in z~1-3 'normal' massive SFGs. Most were drawn from the CO 3-2 observations of z~1-2 SFGs of the Tacconi et al. (2010) and Tacconi & Combes IRAM Large Programs (henceforth 'LP'). The LP data and the smaller CO 2-1 z~1.5 sample of Daddi et al. (2010a) were taken with the IRAM Plateau de Bure millimeter Interferometer (Guilloteau et al. 1992). The LP SFGs are drawn from two samples with median redshifts of <z>=1.2 and <z>=2.2. They are matched to cover the same ranges of stellar mass ($M_*$=3-30 x$10^{10}$ $M_\odot$) and star formation (20-300 $M_\odot$yr$^{-1}$). The LP sample used in this paper has 21 detections between z=1 and 1.5, and 11 detections and 5 upper limits between z=2 and 2.4. The 4 detections from Daddi et al. (2010a) at <z>=1.5 have comparable selection criteria as in the LP sample. To these sets we add the detections of three somewhat lower mass ($M_*$=5-30 x$10^9$ $M_\odot$), strongly lensed SFGs (cB58: z=2.7, Baker et al. 2004, 'cosmic eye": z=3.1, Coppin et al. 2007, 'eyelash': z=2.3, Swinbank et al. 2010, Danielson et al. 2010). For a description of the observations and the data analysis we refer to the papers above.

The galaxies we are analyzing exhibit a correlation between stellar mass ($M_*$) and star formation rate (SFR), or stellar mass and specific star formation rate (sSFR=SFR/$M_*$), the so called 'star formation main-sequence' (Schmininovich et al. 2007, Noeske et al. 2007). Galaxies near the 'main sequence' make up 90% of the cosmic star formation density at z~1.5-2.5 (Rodighiero et al. 2011). The relation has an rms scatter of ±0.3 dex at z~0.5-2 (Figure 1, Elbaz et al. 2007, Noeske et al. 2007, Daddi et al. 2007, Rodighiero et al. 2010, Mancini et al. 2011). 'Main sequence' SFGs have disk-like morphologies with low Sersic indices ($n_S \sim 1$) and, compared to off-main sequence systems, have relatively large effective radii (Wuyts et al. 2011). Most of the galaxies in our sample are extended rotating disks in Hα integral field spectroscopy data sets, HST rest frame UV/optical images, or CO high resolution interferometry maps (Förster Schreiber et al. 2009, Law et al. 2009, Tacconi et al. 2010, Daddi et al. 2010a, Mancini et al. 2011, Combes et al. in prep.). A few are compact, dispersion dominated systems (Law et al. 2009, Förster Schreiber et al. 2009, Tacconi et al. 2010). Their star formation rates thus are 5 to 30 times smaller than bright submillimeter selected galaxies (SMGs) in the same redshift range. SMGs are starbursts above the main sequence and often appear to be major mergers (Greve et al. 2005, Engel et al. 2010). Based on their CO or Hα kinematics, and/or their HST high resolution morphology only one galaxy in our current sample (BX528) is a merger (Tacconi et al. 2010, and in prep., Shapiro et al. 2008, Förster Schreiber et al. 2009).

## 2.2 Metallicities

For 14 $z \geq 1$ SFGs we have individual determinations of gas phase metallicities based on the [NII]/Hα estimator of Pettini & Pagel (2004: $Z_O = 8.9 + 0.57 \log (F([NII])/F(H\alpha))$). Of these 14, 6 galaxies were observed with the SINFONI integral field spectrometer as part of the SINS program (Förster Schreiber et al. 2009), and eight galaxies with long slit spectroscopy (Teplitz et al. 2000, Erb et al. 2006b and priv.comm, Richard et al. 2011, Buschkamp et al. 2011, in prep.). The rms dispersion of the Pettini & Pagel relation is

±0.18 dex for 7.5<$Z_O$<8.6. For the rest we determined metallicities from the stellar mass-metallicity relation at the respective redshifts (Erb et al. 2006b, Buschkamp et al. in prep., Shapley et al. 2005, Liu et al. 2008, $Z_O$ = a + 2.18 log($M_*$) -0.0896 log($M_*$)$^2$, with a=-4.51 for z=1.5-3, and a=-4.45 for z~1-1.5). This includes two AGNs with measured [NII]/Hα (Erb et al. 2006b), for which we also adopt the metallicities estimated from the mass-metallicity relation. The rms dispersion of the z~2 mass-metallicity relation is ±0.09 dex. Metallicities derived from the mass-metallicity relation should thus not be much more uncertain than those inferred from [NII]/Hα. The [NII]/Hα ratio is known (e.g., Pettini & Pagel 2004) to saturate above solar metallicity ($Z_\odot$~8.69 (±0.05), Asplund et al. 2009), and systematic uncertainties between different metallicity indicators and calibrations can exceed 0.3 dex (e.g., Kewley & Ellison 2008). To minimize this systematic effect, we converted all metallicities to the Denicolo, Terlevich & Terlevich (2002) calibration system (also based on [NII]/Hα), with the conversion function given in Table 3 of Kewley & Ellison (2008). The transformation onto the Denicolo et al. scale delivers the best agreement (i.e. smallest scatter) between different metallicity calibrators. It optimizes the comparison to the z~0 metallicity estimates, especially at the high $M_*$ end (Kewley & Ellison 2008), which is particularly important for our study. The systematic uncertainties within the Denicolo et al. (2002) system and over the observed range should be within ±0.2 dex (Kewley & Ellison 2008). Relative uncertainties from the measurement errors in [NII]/Hα and the mass-metallicity relation are much smaller (see typical red error bar at the bottom of Figure 3). For the 14 SFGs with metallicity estimates from both methods the rms scatter between the two methods is ±0.1 dex.

Our final sample of 44 z≥1 SFGs covers inferred oxygen abundances from $Z_O \sim 8.4$ to 8.9 on the Denicolo et al. (2002) scale. For comparison to z~0 SFGs of different metallicities we used the recent compilations of Leroy et al. (2011) and Krumholz et al. (2011). Wherever possible, we replaced their quoted metallicities by [NII]/Hα-based metallicities from the published literature, with the same calibrations as for the high-z data. Most of the [NII]/Hα-derived metallicities are very similar to the ones given by Leroy et al. and Krumholz et al.

## 3. Results

The following analysis was triggered by our finding (initially discussed in Tacconi et al. 2008) an increasing number of galaxies in the IRAM LP z~2 sample that are very faint or not detected in CO 3-2, despite their large star formation rates. Given the PdBI sensitivity and the long (up to 20-30 hours) integration times, we had expected to detect these targets easily. Yet, other galaxies with similar masses and star formation rates were detected as expected. As an example Figure 2 shows the images and normalized Hα and CO 3-2 spectra for two such cases, ZC406690 and Q2343 BX610. These two z~2 SFGs are at opposite extremes of the metallicity distribution of our sample, but otherwise have very similar star formation rates (210-290 $M_\odot yr^{-1}$), dynamical masses ($\geq 10^{11}$ $M_\odot$), sizes ($R_{1/2}$~4-6 kpc), as well as matter surface densities ($10^{2.5-3}$ $M_\odot pc^{-2}$, Genzel et al. 2008, 2011). Both are clumpy rotating disks ($v_{c, max}$~300 km/s). Their stellar masses and metallicities differ by factors of 4 and 2.3, respectively. Their CO 3-2 to extinction corrected Hα flux ratios differ by 5.4. While the CO 3-2 line is well detected in BX610 ($Z_O$=8.8) it is at best marginally detected in ZC406690 ($Z_O$=8.4). Likewise in the clumpy, rotating disk BX482 ($Z_O$=8.5, SFR~120 $M_\odot yr^{-1}$, $R_{1/2}$=4.2 kpc, $v_{c\,max}$=240 km/s, Genzel et al. 2008, 2011, Förster Schreiber et al. 2009, 2011) the ratio of the 3σ upper limit in CO 3-2 to the extinction corrected Hα flux is 3 times lower than in BX610.

As we now discuss, we conclude that these puzzling findings are probably not random but are caused by the dependence of the galaxy integrated CO luminosity per molecular gas mass on gas phase oxygen abundance. To prove convincingly that such an effect is present requires an estimate of the intrinsic molecular ($H_2$) gas mass. Unfortunately this cannot be obtained directly. We argue in the next sub-section that the high-z Kennicutt-Schmidt relation between molecular gas mass and star formation rate is now well enough understood that it can be used to derive a good enough measure, albeit indirect, of the intrinsic molecular gas mass of a distant galaxy.

### 3.1 The Kennicutt-Schmidt (KS) relation

In his seminal 1998 paper Kennicutt (1998, henceforth K98) established that the star formation surface density $\Sigma_{star\,form}$ and the total (atomic + molecular) gas surface density $\Sigma_{gas}$ in z~0 star forming galaxies (including ultra-luminous infrared galaxy mergers (ULIRGs)) are related through

$$\log(\Sigma_{star\,form}) = a + n \times \log(\Sigma_{gas}) \qquad (3),$$

with a best fit slope of n~1.4. More recent studies of galaxy integrated and spatially resolved KS-relations (on kpc scales) in non-merger SFGs near the star formation main-sequence and near solar metallicity have found that the star formation rate surface density correlates mainly with the molecular gas surface density, $\Sigma_{mol\,gas}$ (Kennicutt et al. 2007, Bigiel et al. 2008, Schruba et al. 2011). The 'molecular' KS-relation is flatter, with n~1.0 to 1.3 for $\Sigma_{mol\,gas}$~10 to $10^{3.5}$ $M_\odot pc^{-2}$ for near-solar metallicity disk galaxies and a constant Galactic conversion factor (Kennicutt et al. 2007, Bigiel et al. 2008, Leroy et al. 2008, Daddi et al. 2010b, Genzel et al. 2010, Schruba et al. 2011).

We show in Figure 3 an update of the galaxy integrated KS-relations shown in Figures 2 and 3 of Genzel et al. (2010), amended with the additional new z≥1 data discussed in this paper. We used $\alpha_{CO\,1-0}=\alpha_G$ for all data. To our knowledge Figure 3 is the largest compilation of both low- and high-z galaxy integrated KS-measurements so far in the literature. The z~0 SFG sample (all observed in CO 1-0) includes many more SFGs, moderate starburst galaxies (e.g. M82, NGC253, NGC3256), and luminous infrared galaxies (LIRGs) than K98, but no ULIRG mergers (see Genzel et al. 2010 for details). These additional z~0 SFGs span more than three orders of magnitude in molecular gas surface density, from the lower limit of the mostly molecular gas dominated regime at 5-10 $M_\odot\,pc^{-2}$ (Blitz & Rosolowsly 2006) to dense starbursts at >$10^4$ $M_\odot\,pc^{-2}$. The z~0 SFGs also overlap with the entire range of the z≥1 SFGs (observed in 3-2 for the z~1-1.5 and z~2-2.4 IRAM galaxies, and the lenses, and in CO 2-1 for the z~1.5 Daddi et al. 2010a SFGs). The spatially resolved KS-studies of Kennicutt et al. (2007), Bigiel et al. (2008), Leroy et al. (2008) and Schruba et al. (2011) cover the range of 10 to 100 $M_\odot pc^{-2}$ and are in excellent agreement with the galaxy integrated data shown here. For computing surface densities, half light radii were taken from molecular observations themselves at

z~0, and from a mixture of CO, Hα- and rest-frame UV-continuum sizes for z>1 (Genzel et al. 2010).

For the best examination of the slope of the KS-relation, we plot in Figure 3 on the vertical axis the ratio of star formation rate (or its surface density) to molecular gas mass (or its surface density), inferred from the CO line observations with $\alpha=\alpha_G$. This is the molecular gas depletion rate $(t_{depletion}(CO))^{-1}$. An n=1 KS-relation (a constant depletion rate independent of gas surface density) corresponds to a horizontal line. The best fit to all z~0 SFGs and the z≥1 SFGs above solar metallicity ($Z_O$>8.7) yields a slope of n=1.1 (±0.06), in the middle of the range of slopes found in the literature from both spatially resolved and galaxy integrated z~0 studies. The slope of the z≥1 SFGs with $Z_O$>8.7 is marginally flatter (n=0.9±0.12). Given the fairly large (±0.3 dex, K98, Kennicutt et al. 2007, Figure 3) scatter of the relation, as well as the systematic uncertainties in the slope determinations of ±0.2 to ±0.25 (see discussion in Genzel et al. 2010), Occam's razor suggests that all data sets are consistent with a universal, near-unity slope, assuming that $\alpha_{CO}$ does not vary systematically with $\Sigma_{mol\ gas}$ (see the discussion at the end of this section).

There does appear to be a small but statistically significant offset in the average depletion rates at z~0 and z>1 (Figure 3 and caption, Genzel et al. 2010). The average depletion rate is 0.4 to 0.5 $Gyr^{-1}$ at z~0 and ~1 $Gyr^{-1}$ at z=1-2.5 (Leroy et al. 2008, Bigiel et al. 2011, Tacconi et al. 2010, Daddi et al. 2010a,b, Genzel et al. 2010, Bauermeister et al. 2010, Saintonge et al. 2011). Saintonge et al. (2011) find a positive correlation of the depletion rate with sSFR, with a ~0.5 dex difference in depletion rate between the highest and smallest sSFR SFGs in Figure 3. This variation is quite consistent with the scatter in Figure 3 (rms ±0.3 dex for z~0 SFGs, and ±0.2 dex for $Z_O$>8.7 z>1 SFGs).

To explore the dependence of the KS-slope on CO excitation and the rotational quantum number J of the observed line, Narayanan et al. (2011a) have carried out numerical hydro-simulations of star formation in the molecular interstellar medium with an input volumetric star formation relation, $\dot{\rho}_{starform} \propto \rho_{mol\,gas}/t_{ff} \propto \rho_{mol\,gas}^m$. Here $t_{ff} \propto \rho^{1-m}$ is the local free fall time in the molecular clouds. Narayanan et al. find that the resulting gas-star formation surface density relation depends on the rotational quantum number J of the CO line used for tracing the molecular gas. For m=1.5 and conditions similar to z≥1 SFGs or z~0 starbursts the simulations yield n(J=1)~1.47 and n(J=3)~1.08. The data in Figure 3 are clearly not consistent with this result, since the z~0 data (observed in CO 1-0) and the z>1 data (observed in CO 2-1 and 3-2) both have n~1 and the difference in slope is less than 0.3, even if the fit uncertainties are included. A better match to the simulations occurs for m=1 ($t_{ff}$=const), for which Narayanan et al. predict n(J=1)=0.96 and n(J=3)=0.76. This case would apply if the molecular interstellar medium consists of a collection of clouds with similar properties (Bigiel et al. 2008), or if the galactic disk is in a state of marginal gravitational stability, such that the Toomre Q-parameter is near unity, and the local free fall time scale above becomes similar to the global dynamical time scale (c.f. Genzel et al. 2010).

Could there be a dependence of $\alpha_{CO}$ on $\Sigma_{mol\,gas}$? There are plausible reasons discussed in section 3.3 and Tacconi et al. (2008) (due to the $n(H_2)^{1/2}/T$ dependence of $\alpha_{CO}$) that such a dependence should be present to some extent (see Appendix A and Figure 10 in Tacconi et al. 2008). Based on the available empirical data the possible change of $\alpha_{CO}$ in the range sampled in Figure 3 may be about a factor of 2 (0.3 dex). To explore the impact of such a change we have applied an artificial shift of -0.3 dex to all galaxies with surface densities >$10^3$ $M_\odot pc^{-2}$ (M82 is at $10^{3.4}$), and then refitted the KS relation. This results in an overall KS-slope of 1.22 instead of 1.1, again within the systematic slope uncertainties.

A final point relates to the issue of atomic hydrogen in the KS-relation. We have mentioned that all galaxies in Figure 3 are above ~5-10 $M_\odot pc^{-2}$ where the z~0 pressure-$H_2$ relation of Blitz & Rosolowsky (2006) indicates that almost all hydrogen is in molecular form. However the combined effects of strong stellar feedback, shocks and turbulence in starbursts and high-z SFGs may lower the fraction of $H_2$ to total (star forming) gas at a given gas surface density, relative to that in normal z~0 disks. In this case the gas surface densities are lower limits to the total (star forming) gas in each galaxy. It is plausible that this correction decreases from low to high gas surface densities, suggesting that the intrinsic KS-relation (with these corrections added) is still flatter than indicates by the data in Figure 3..

## 3.2 The ratio SFR/($α_G$ $L_{CO\ 1-0}$) increases with decreasing metallicity

Given this fairly universal KS-relation for massive SFGs at both low- and high-redshifts, we can now ask whether there is a systematic dependence of the depletion rate on metallicity. In Figure 3 we plot the depletion time scale of the z>1 SFGs as a function of $\Sigma_{mol\ gas}$ for three different metallicity bins, assuming $α_G$=4.36 for all bins. Super-solar, high metallicity galaxies ($Z_O$> 8.76) are marked by blue filled circles, SFGs with 8.76>$Z_O$>8.62 are denoted by black triangles, and the lowest metallicity galaxies in our sample ($Z_O$<8.62) are marked by filled red squares. There is a significant difference in the inferred depletion rates for these three metallicity bins. At least for the highest and middle bins the number of galaxies is large enough to see that the galaxies in each bin follow a fairly flat KS-relation but are offset from one another. The offset in the lowest metallicity bin is even larger but the number of galaxies is too small to estimate the KS-slope. This appearance of flat KS-relations 'peeling off' vertically as a function of $Z_O$ is exactly what is expected if the variations were due to differences in $\alpha_{CO}(Z_O)$ between the metallicity bins.

Another representation of the same data is given in Figure 4, where we plot the depletion rate as a function of gas phase oxygen abundance in the z~1-2.5 SFGs of our sample. At or above solar metallicity all z≥1 SFGs approach SFR/$\alpha_G L'_{CO\ 1-0}$ ~1 Gyr$^{-1}$, in agreement with the discussion above, and with a scatter that is consistent with the measurement uncertainties. Below solar metallicity, the data exhibit a trend of rapidly increasing SFR/$\alpha_G L'_{CO\ 1-0}$ with decreasing oxygen abundance. The trend at z≥1 is similar to that found at z~0. Data points with metallicities derived from [NII]/Hα and from the mass-metallicity relation agree well but the overall scatter is quite large.

## 3.3 The variation in SFR/$L_{CO\ 1-0}$ is not due to changes in physical gas depletion time or ISM conditions

Can the trends in Figures 3 and 4 be driven by a physical change in depletion rate or in the ISM properties, and is metallicity the primary underlying variable? The lowest metallicity z~0 star-forming systems are dwarf/irregular galaxies, such as the SMC and NGC 6822. Given the evidence for time variable star formation histories in such systems (Tolstoy, Hill & Tosi 2009), their much greater depletion rates compared to normal disk galaxies might be the result of recent short-duration starbursts. However, combined spatially resolved studies of HI, infrared dust and CO emission in the SMC and a number of the other z~0 SFGs plotted in Figure 4 – that yield the CO conversion factor and the molecular gas mass without relying on the KS-relation - strongly suggest that it is the absence of CO emission, and not the presence of starburst events that dominate the apparently high depletion rates (Leroy et al. 2011, Bolatto et al. 2011).

With the exception of cB58, the starburst explanation is even more unlikely for the high-z SFGs in our sample. Almost all are massive systems on or near the star formation main-sequence (Figure 1). Galaxies near the main sequence exhibit exponential light profiles with fairly large ($R_{1/2}$~3-6 kpc) disk radii (Wuyts et al. 2011). They exhibit high star formation duty cycles (30-100 %, Adelberger et al. 2005, Noeske et al. 2007, Daddi et al. 2007). High-z main-sequence SFGs are forming stars at high rates (20-300 $M_\odot yr^{-1}$) primarily because of their high gas fractions (Daddi et al. 2010a, Tacconi et al. 2010) and not because they are undergoing short-duration 'starburst' events. The high gas fractions are plausibly driven by the semi-continuous large gas accretion rates predicted by all cold dark matter models (e.g. Kereš et al. 2005, Dekel & Birnboim 2006, Ocvirk, Pichon & Teyssier 2008, Genel et al. 2008, Bouché et al. 2010). At fixed redshift the star formation surface densities are almost constant as a function of stellar mass (and thus metallicity) along the main-sequence (~0.6 and ~3 $M_\odot yr^{-1} kpc^{-2}$ for z~1 and 2, Wuyts et al. 2011).

With the possible exception of the lensed 'eyelash' galaxy (Swinbank et al. 2010) all galaxies in our sample are within the ±0.3 dex dispersion of the $M_*$-sSFR main sequence relations at z~1.2 and 2.2, including the individual uncertainties in stellar masses and star formation rates (Figure 1). Merger induced starbursts (e.g. z~0 ULIRGs) with internally enhanced depletion rates typically lie an order of magnitude or more above the main-sequence line (Daddi et al. 2010b, Genzel et al. 2010, Combes et al. 2011, Saintonge et al. 2011).

Recent Herschel PACS observations have revealed a remarkable uniformity of the infrared spectral energy distributions of massive main-sequence SFGs at all redshifts (Hwang et al. 2010, Elbaz et al. 2011, Nordon et al. 2011). Main-sequence galaxies with star formation rates from a few to a few hundred solar masses per year (with the exception of z~0 ULIRGs) have similar dust temperatures, $T_{dust}$~27-38 K, between z~0 and 2 (Hwang et al. 2010). High-z SFGs are merely 2-5 K colder than z~0 SFGs of the same luminosity.

Observations of multiple CO rotational lines in a number of z≥1 SFGs and submillimeter galaxies show that the CO ladder distributions are similar to those of local starburst galaxies, such as M82 and NGC253, with inferred local molecular hydrogen volume densities (for the CO 3-2 emission) of $n_{H2}=10^{2.5...4.5}$ cm$^{-3}$ (Weiss et al. 2007, Dannerbauer et al. 2009, Danielson et al. 2010, Riechers et al. 2011, Combes et al. in prep.). Average molecular hydrogen densities in the giant star forming clumps in z≥1 SFGs may be ~$10^{2..3}$ cm$^{-3}$ (Genzel et al. 2011). However, the fortuitous cancellation of the density and temperature dependencies may break down very close to massive star formation sites and may drive the conversion factor downward, as in the case of the central starburst region in M82 ($\alpha_{CO\ 1-0}$ ~2, Wild et al. 1992, Weiss et al. 2001). This means that the average ratio $\frac{(<n(H_2)>)^{1/2}}{T_{R\ 1}}$ in equation (2) probably is comparable in

z≥1 SFGs to that in moderate z~0 starbursts, or the Milky Way GMC population ($T_{R\,1}$~ 7-30 K, $<n_{H2}>$~$10^{1.5...2}$ cm$^{-3}$).

Could the large turbulence in high-z SFGs affect the observed metallicity trend in $\alpha_{CO\,1-0}$? As a rule z≥1 SFGs near the 'main sequence' exhibit 4 to 10 times larger velocity dispersions than z~0 SFGs (Förster Schreiber et al. 2009, Law et al. 2009, Epinat et al. 2009). Recent theoretical work suggests that increased turbulence may profoundly affect the local density and temperature structure, and in turn also the conversion factor (Narayanan et al. 2011b, Shetty et al. 2011a,b). However, the observed velocity dispersions in z≥1 main-sequence SFGs appear to depend little on galaxy mass, star formation rate or surface density (Genzel et al. 2011 and references therein). In addition, for most of the z≥1 SFGs in Figures 1 and 4 rotation dominates over random motions, in contrast to many z~0 ULIRGs and z≥1 submillimeter galaxies (Tacconi et al. 2008, Engel et al. 2010). Star formation in z≥1 main sequence SFGs is plausibly driven by gravitational instabilities in giant star forming clouds, similar to GMCs but scaled-up to the large gas fractions at z≥1 (Genzel et al. 2011). In any case the effect of turbulent compression and the presence of non-virialized gas components with chaotic motions drive the conversion factor downward, not upward, both in the empirical data (Solomon et al. 1997, Scoville et al. 1997, Downes & Solomon 1998, Tacconi et al. 1999, 2008), as well as in the simulations of Narayanan et al. (2011b) and Shetty et al. (2011b).

Another issue is whether the CO rotational ladder excitation might vary between galaxies in a systematic way such that a constant brightness temperature scaling factor $R_{1J}$ from J to 1, as used in this study, is not appropriate. The first studies of the CO rotational ladder distributions indeed show variations in CO rotational excitations but for J≤3 these are by far too small to account for the magnitude of the variations in Figure 4 (Weiss et al. 2007, Dannerbauer et al. 2009, Danielson et al. 2010, Riechers et al. 2011, Ivison et al. 2011, Combes et al. in prep.).

## 3.4 The variation in SFR/$L_{CO\,1-0}$ cannot be explained by the fundamental-metallicity relation

Mannucci et al. (2010) have shown that the gas phase oxygen abundance in z~0 SFGs not only depends on stellar mass (i.e. the mass-metallicity relation discussed in section 2.2) but also on star formation rate. At fixed stellar mass a galaxy with higher star formation rate has a lower oxygen abundance,

$$12 + \log(O/H) = b + \beta \times \left(\log(M_*) - \gamma \log(SFR)\right) \quad (4),$$

where b=4.2, β=0.47 and γ=0.32 for metallicities on the Maiolino et al. (2008) scale (Mannucci et al. 2010). As we discuss in this section this 'fundamental metallicity relation' (FMR) in combination with a non-linear (n>1) KS-relation may in principle yield a depletion rate – metallicity relation as in Figure 4.

For z~0 galaxies in the Sloan Digital Sky Survey the FMR has an rms scatter of ±0.05 to ±0.06 dex, which is smaller than the ±0.08 dex scatter in the mass-metallicity relation (Mannucci et al. 2011), suggesting it may be the more fundamental relation. Mannucci et al. (2010) interpret the FMR as being driven by the interplay of star formation increasing metallicity and accretion delivering fresh, low metallicity gas. Mannucci et al. (2010)

also show that the 10 z~1-2.5 SFGs with reliable metallicity estimates available to their study fall on the same relation, indicating that equation (4) may also apply at z~1-2.5. If the (molecular) KS-relation is non-linear (n>1 in equation 3), the combination of equations (3) and (4) yield

$$t^{-1}_{depletion}(\text{mol gas}) \propto M_*^{\delta} \times 10^{-\chi Z_o} \times R_{1/2}^{-(2n-2)/n}, \text{ with}$$

$$\delta = (1-1/n)/\gamma \text{ and } \chi = (1-1/n)/(\beta\gamma) \quad (5).$$

For n=1.1 and the FMR parameters found by Mannucci et al. (2010: β=0.47, γ=0.32) δ= 0.28 and χ=0.6. For a fixed stellar mass equation (5) then predicts that the depletion rate increases by a factor of 2 between solar metallicity and 0.5 times solar metallicity. This is in qualitative agreement with the observations shown in Figure 4. The FMR predicts that at fixed stellar mass a lower metallicity galaxy has a higher star formation rate. If the KS-relation is non-linear, a higher star formation rate automatically implies a greater depletion rate. But, can equation (5) account also quantitatively for the observations shown in Figure 4?

The average amplitude of the FMR is already captured in our analysis by using a redshift dependent mass-metallicity relation, as described in section 2.2. Additional changes in metallicity estimates per se from the scatter of the stellar mass - star formation rate relation (±0.3 dex in SFR at fixed stellar mass) are smaller (<0.1 dex) than other uncertainties. To estimate the best FMR parameters at z~2 we used a recent, yet unpublished compilation of z~1.4-2.5 SFGs observed in [NII] and Hα with integral field long-slit spectroscopy as part of the SINS, OSIRIS and LUCIFER surveys (Förster

Schreiber et al. 2009, Law et al. 2009, Buschkamp et al. in preparation). The best fit z~2 FMR on the Denicolo et al. (2002) scale has β=0.31 (±0.06 1σ) for data stacked in stellar mass bins (all for γ fixed at 0.32). The rms scatter of the z~2 data is ±0.11dex.

We list in Table 1 the molecular depletion rates and star formation rates computed from equation (5) for the four most extreme SFGs in Figure 4, and compare these values to the corresponding values computed from equations (3) and (4), for n=1.1 and 1.3, and for β=0.31, 0.25 and 0.37 (reflecting the 1σ uncertainty in the z~2 FMR relation). The value of β=0.37 on the Denicolo et al. (2002) scale is equivalent to β=0.47 on the Maiolino et al. (2008) scaled deduced by Mannucci et al. (2010) to be the best fit the z~0 SFGs. With this range of values Table 1 shows that the effect of the FMR cannot simultaneously account quantitatively for the observed star formation rates and depletion rates (for constant $\alpha_G$=4.36) in ZC406690, cB58 and BX389, for any combination of the parameters n and β. Either the predicted depletion rate is too small, or the star formation rate is too large, or both. Obviously no effect is predicted for a KS slope of n=1, favored by the work of Bigiel, Leroy, Schruba and collaborators. The largest impact of the FMR naturally is for steeper KS-slopes (n=1.3), as favored by Kennicutt et al. (2007) and Daddi et al. (2010b). In the case of BX482 a match is possible for small β~0.2-0.3 and n>1.1.

The unrealistically large star formation rates found for many of the cases in Table 1 immediately show that the strict application of an 'inverse' FMR (as in equation (5)) is not a good tool for individual galaxies. This is because the dependence on star formation in equation (4) is very shallow (β×γ~0.1) and the scatter of the FMR is large. This means that the inverse solution for star formation rate depends very strongly on the location of a source in the $Z_O$-$M_*$ plane and the assumption of a scatter free relation.

We conclude that including the effect of the FMR may indeed somewhat decrease the amplitude of the effect seen in Figure 4. However, the FMR cannot account for the observed increase of depletion rate with decreasing metallicity.

## 3.5 The variation in SFR/$L_{CO\ 1-0}$ may be caused by photodissociation of CO in low-metallicity environments

Theoretical work on UV-illuminated molecular clouds reaching back 25 years has consistently predicted a strong dependence of $(t_{depletion}(CO))^{-1}$ on metallicity, especially in somewhat diffuse molecular gas (van Dishoeck & Black 1986, 1988, Maloney & Black 1988, Sternberg & Dalgarno 1995, Hollenbach & Tielens 1999, Wolfire et al. 1990, 2010, Bolatto, Jackson & Ingalls 1999, Papadopoulos & Pelupessy 2010, Shetty et al. 2011a, Glover & Mac Low 2011, Glover & Clark 2011). The short-dashed blue curve in Figure 4 (from Krumholz et al. 2011) is the result of calculating $(t_{depletion}(CO, Z_O))^{-1}$ for a spherical molecular cloud with constant column density ($\Sigma_{gas}\sim 80\ M_\odot pc^{-2} \sim <\Sigma_{gas}(GMC, MW)>$), and exposed to a diffuse UV radiation field with a characteristic ratio of UV energy density $G_{UV}$ to gas density $n_H$ similar to that in the solar neighborhood ($G_{sn}=2.7\times 10^{-3}$ erg cm$^{-2}$ s$^{-1}$, $n_{sn}=23$ cm$^{-3}$, Wolfire et al. 2010, Krumholz et al. 2011). For this curve the volume filling factor $f_V$ (the inverse of the clumping factor c as defined by Krumholz et al. 2011) of molecular gas is assumed to be 0.2. This curve represents the ratio of the total $H_2$ column through the cloud to the $H_2$ column in which CO remains molecular, given the adopted UV radiation field and its density, clumpiness and total column (Wolfire et al. 2010). The depth of the CO-photodissociation zone is controlled by dust ($A_{UV} \sim 1$) and thus scales directly with metallicity (e.g. van Dishoeck & Black 1986, 1988, Maloney & Black 1988, Sternberg & Dalgarno 1995, Wolfire et al. 2010, Glover & Mac Low 2011, Feldmann et al. 2011). The work of Krumholz et al. (2011) and Glover & Clark (2011) implies that star formation does not require the formation/presence of CO but can also occur in 'CO-dark' gas. Krumholz et al. (2011) find that this star forming 'CO-dark' gas is molecular (in terms of hydrogen), while Glover & Clark (2011) conclude that star formation can even occur in atomic hydrogen gas as long as dust shielding is efficient ($A_V > 3$).

Simple estimates based on star formation rates and galaxy sizes suggest that $z \geq 1$ SFGs in Figure 1 have UV radiation field densities $\sim 10^{2...3}$ times that in the solar neighborhood. Average ISM densities may also be greater, such that the ratio is probably similar to that in the solar neighborhood, or in nearby starburst galaxies. The onset of the up-turn in the blue-dashed curve in Figure 4 depends strongly on $f_V$. Model curves with larger $f_V$ values than chosen by Krumholz et al. (2011) have up-turns at higher metallicity. For a homogeneous cloud ($f_V=1$) the up-turn occurs at solar metallicity.

The agreement of the theoretical predictions with the observations suggests that the trends seen in Figure 4, at both low and high redshifts, may mainly be the consequence of the ratio of the volume and mass of molecular gas traced by CO (relative to that in $H_2$) decreasing rapidly at low metallicity due to photodissociation by the ambient UV-field (see Pelupessy & Papadopoulos 2009, Krumholz et al. 2011, Glover & Mac Low 2011, Shetty et al. 2011b).

## 3.6  An empirical scaling relation for $\alpha_{CO}(Z_O)$

We now turn the results in Figure 4 around and derive an empirical dependence of $\alpha_{CO\ 1-0}$ on metallicity, based on the assumption of a universal KS-scaling relation. Using the KS-relation in reverse, the molecular gas mass can be obtained from the star formation rate (e.g. Erb et al. 2006b),

$$M_{mol\ gas}(M_\odot) = \xi\ SFR(M_\odot yr^{-1})^{1/n}\ R_{1/2}(kpc)^{(2n-2)/n} \qquad (6).$$

$R_{1/2}$ is the half-light radius of the gas/star forming disk. For the z>1 data with $Z_O$>8.7 in Figure 3 and the best fit slope of n=1.1 we find $\xi$=1.02x10$^9$. For a slope fixed to n=1.3 (Daddi et al. 2010b), the data in Figure 3 yield $\xi$=1.2x10$^9$. In comparison to equation (6) the simpler assumption of a constant depletion rate of 1 Gyr$^{-1}$ (n=1) at z>1 yields typically 5% larger gas masses for all but the most compact galaxies. The relation given in equation (8) of Genzel et al. (2010), depending on the ratio of $R_{1/2}$ to the circular velocity (i.e. the global dynamical time), instead of $R_{1/2}$ alone, on average yields half the masses. The relation in Kennicutt et al. (2007, n=1.37) yields 20% smaller gas masses. These differences are all within the scatter of the empirical KS-relation.

We apply equation (6) to each galaxy and divide the inferred gas mass by $L'_{CO}$. The result is a direct measure of the conversion factor, and Figure 5 shows the derived CO 1-0 conversion factor as a function of metallicity for our z≥1 sample. Figure 5 has two insets. The left inset applies for our favored KS-slope of n=1.1 (Figure 3). The right inset is for

n=1.3 favored by Daddi et al. (2010b) and Kennicutt et al. (2007). For comparison, we show estimates of $\alpha_{CO\,1-0}$ of z~0 SFGs by Leroy et al. (2011). The Leroy et al. (2011) estimates of $\alpha_{CO\,1-0}$ come from simultaneous parametric fitting of spatially resolved HI, CO and dust emission. They do not rely on the KS-relation. The three sets of dashed lines denote previous CO conversion factor scaling relations for z~0 SFGs (Wilson 1995, Arimoto et al. 1996, Israel 1997, 2000, Boselli et al. 2002). This earlier work was partially based on similar galaxies as in the Leroy et al. (2011) sample but employed different methods.

In agreement with the theoretical expectations, the CO → $H_2$ conversion factor increases with decreasing metallicity for all these samples, with similar trends at both low- and high-z, albeit with a large scatter. If the z~0 points of Leroy et al. (2011) and the z≥1 SFGs are combined (treating 3σ upper limits as detections), the best linear fit yields the relation

$$\log(\alpha_{CO\,1-0}) = -1.3\,(\pm 0.25)\,(12 + \log(O/H))_{\text{Denicolo 02}} + 12\,(\pm 2) \quad (7),$$

where the quoted uncertainties are 2σ fit uncertainties for equal weights to all data points (because of the dominance of the systematic errors discussed above). The result in equation (7) is the same for n=1.1 and n=1.3. For n=1.1 a fit to only the z≥1 data yields a slope of -1.8 (±0.6) and zero point of 17 (±6), which is somewhat steeper than, but not significantly different from the combined fit. This slope may get somewhat flatter, however, depending on the FMR effects and the slope of the KS-relation, as discussed in section 3.4. Our method constrains the relative conversion factor-metallicity relation to no better than ±50%, since the rms scatter of the data points around any of the two relations given above is ±0.23 dex. Future observations are needed to clarify whether the outliers in Figure 4 are due to observational uncertainties, or whether they might point to intrinsic variations in the conversion factor, perhaps as a result of variations in the clumping factor. Absolute uncertainties are larger because of the inherent uncertainties in the metallicity calibrations (and their applicability at high-z), stellar masses and star formation rates and the excitation of the CO rotational ladder. These uncertainties reflect a combination of the systematic uncertainties and plausible physical variations.

## 4. Conclusions

We have analyzed the empirical evidence for a metallicity dependence of the CO luminosity to molecular gas mass conversion factor $\alpha_{CO\ 1-0}$, based for the first time on both low- and high-redshift star forming galaxies. We find that the CO-based molecular gas mass depletion rate in massive z>1 star forming galaxies increases with decreasing gas phase metallicity estimated from strong rest-frame optical emission line ratios or the mass-metallicity relation. We interpret this trend as being mainly driven by the dependence on metallicity of the ratio of galaxy averaged, gas column traced by CO emission to the total $H_2$ column, consistent with the expectations from photodissociation theory. Very similar trends are seen in z~0 star forming galaxies. If correct our findings imply that stars are able to form efficiently in 'CO-dark' gas, as proposed in the recent theoretical work of Krumholz et al. (2011) and Glover & Clark (2011).

We then employed the KS- relation for high metallicity, near-main sequence SFGs at z≥1 to derive empirical CO conversion factors. Combining our sample of 44 z≥1 SFGs with KS-independent conversion factors derived for 11 z~0 SFGs from the compilation by Leroy et al. (2011) we find that the $\log(\alpha_{CO\,1-0})$ - $(12 + \log\{O/H\})$ relation has a slope between -1 and -2. There may be a tendency that at z≥1 the conversion factor tends to decrease non-linearly with metallicity, almost as the square of $Z_O$, while at z~0 the dependence is still compatible with a linear relation. The more pronounced non-linear dependence might be due to the larger star formation rates at high redshift. At 0.5 (respectively 0.25) times solar metallicity $\alpha_{CO\,1-0}$ is ~2.5 to 4 times (respectively 6 to 14 times) larger than at solar metallicity. The uncertainties of the inferred $\alpha_{CO\,1-0}$ values are ±0.23 dex statistically and larger systematically, and are driven by the large measurement and calibration uncertainties, our small galaxy samples and potentially additional 'hidden' parameters and dependencies. Because of the obvious importance of the functional dependence of the CO conversion factor on metallicity and ISM parameters for future large molecular gas surveys it is highly desirable to improve the statistical robustness and uncertainties of the present result by enlarging the samples and their parameter ranges, in order to be able to marginalize over these other parameters.

The implications of our findings may be particularly relevant for redshifts near and above the peak of cosmic star formation activity (z~1-2.5). Because of the cosmic evolution of the mass-metallicity relation a galaxy at the knee of the stellar mass function ($M_S$~$10^{11}$ $M_\odot$) has ~0.74, 0.69 and 0.55 solar metallicity at z~1, 2.2 and 3.5 (Maiolino et al. 2008). A 0.1 $M_S$ galaxy has typically half of the metallicity of an $M_S$ galaxy. These numbers immediately show that CO-based gas mass measurements may need to be significantly revised upwards at z>1 even for 0.7 $M_S$ galaxies, and ≤0.1 $M_S$ galaxies at z>2 may become hard to detect even with the superior sensitivity of ALMA.

*Acknowledgments.* We thank the anonymous referee for valuable comments that have improved the paper. AS thanks the DFG for support via German-Israeli Cooperation grant STE1869/1-1.GE625/15-1.

# Figures

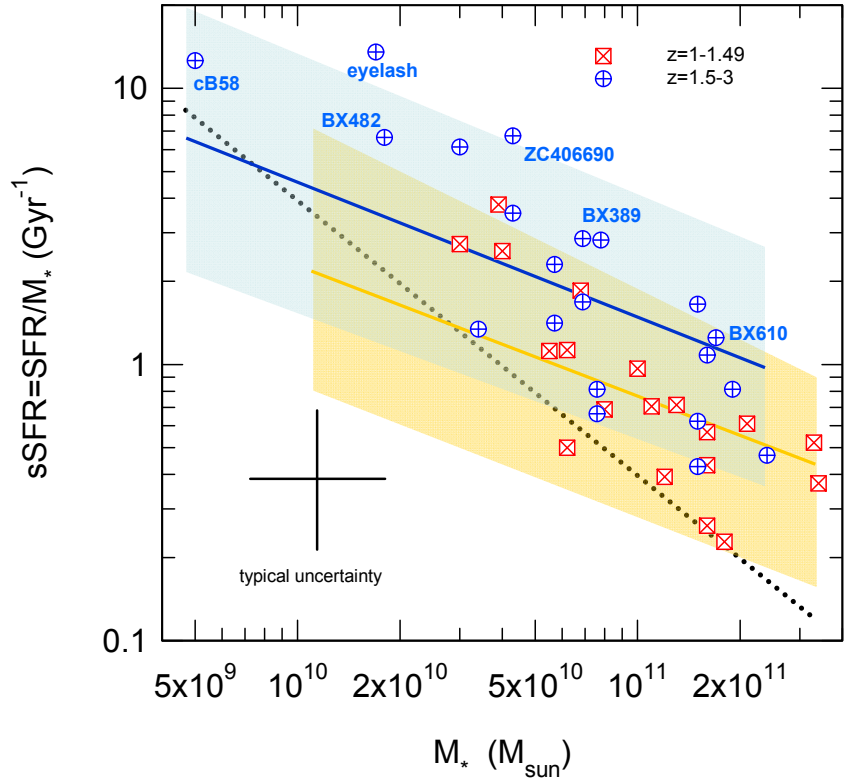

Figure 1. Specific star formation rate (sSFR=SFR/M∗) as a function of stellar mass for the $z=1-1.4$ SFGs (open crossed red squares) and $z=1.5-3$ SFGs (open crossed blue circles) of our sample. The orange and blue shaded regions (and orange and blue lines) denote the location of the 'main-sequence' at these redshifts, as determined from Noeske et al. (2007), Rodighiero et al. (2010), Förster Schreiber et al. (2009) and Mancini et al. (2011). The dotted line denotes the LP survey limit in star formation rate (SFR~ 40 $M_\odot yr^{-1}$).

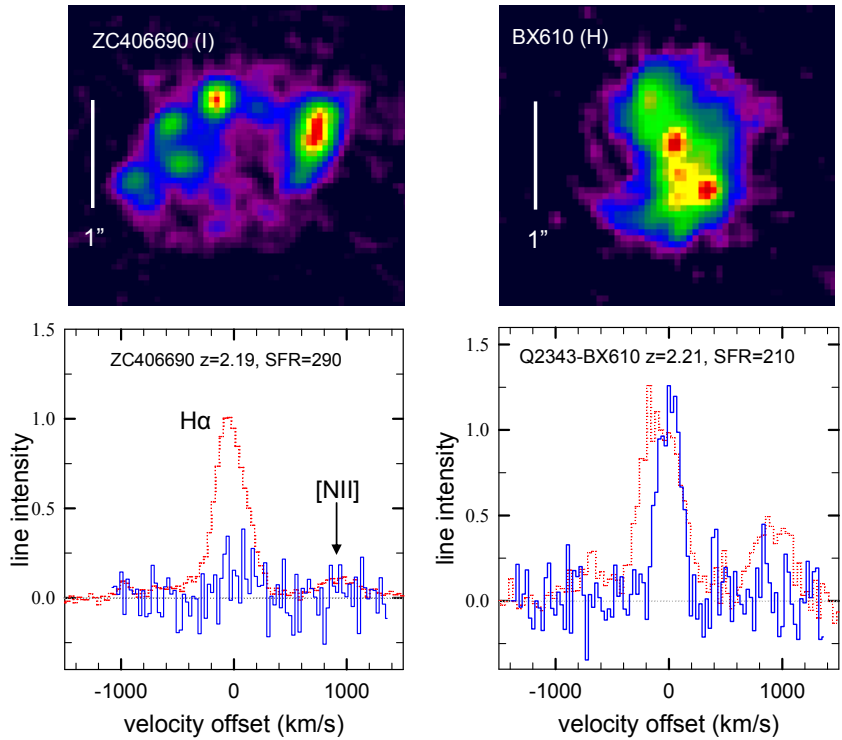

Figure 2. Variation in CO flux to Hα flux for two massive z~2 SFGs with different metallicities (and stellar masses), at the extremes of the observed distribution, but otherwise similar star formation rates, sizes and dynamical masses. The top and bottom insets show the rest-frame UV/optical stellar images (Genzel et al. 2011, Förster Schreiber et al. 2011), Hα/[NII] (dotted red, Genzel et al. 2011, Förster Schreiber et al. 2009) and CO 3-2 spectra (continuous blue, this paper) for ZC406690 (Mancini et al. 2011) and Q2343 BX610 (Erb et al. 2006b, Genzel et al. 2008). The CO 3-2 and Hα/[NII] spectra are normalized such that they both have a peak normalized value of unity for BX610. The low metallicity SFG ZC406690 has 5.4 times smaller CO to Hα flux ratio than the high metallicity SFG BX610.

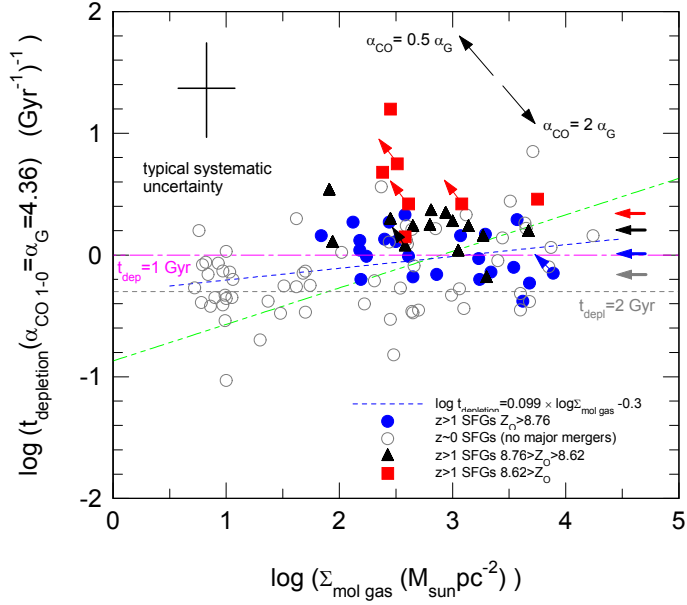

Figure 3. CO-based molecular gas depletion rate ( $(t_{depletion})^{-1}$ = SFR/($\alpha_G$ L'$_{CO1-0}$)) as a function of molecular gas ($H_2$ + He) surface density. The molecular gas mass is inferred from the observed CO 3-2/2-1 flux/luminosity from equations (1) and (2) with $R_{13}$=2 (1.5 for the eyelash) or $R_{12}$=1.16, and $\alpha_{CO\ 1-0}$=$\alpha_G$=4.36. The open grey circles are non-merger, near main-sequence and moderately high mass z~0 SFGs from the compilation in Genzel et al. (2010). The z~0 sample does include many more LIRGs and moderate starbursts than the K98 sample, but no ULIRG mergers. Blue circles, black triangles and red squares denote z>1 SFGs from the IRAM LP program (Tacconi et al. 2010, and unpublished), plus the z~1.5 SFGs from Daddi et al. (2010a), in the metallicity ranges $Z_O$=12+log(O/H) >8.76, 8.62 to 8.76 and <8.62, respectively. Oxygen abundances $Z_O$ for the z>1 SFGs were derived either from the [NII]/Hα ratio, or from the stellar mass-metallicity relation, as described in section 2.2. The metallicities are on the Denicolo et al. (2002) calibration. In this form of the KS-relation constant molecular depletion times

or rates are horizontal lines. Depletion time scales of 1 and 2 Gyrs are marked by dashed grey and magenta lines. The dotted blue line is the best overall linear fit (KS-slope n=1.1) to all z~0 SFGs, plus the super-solar ($Z_O$>8.7) tail of the z>1 SFGs. The long-short dash green line is the best fit of slope 0.3 (KS-slope n=1.3). The different colored arrows on the right side denote the mean depletion rates for the galaxy samples of the same color. These are -0.16 (±0.04) for 62 z~0 SFGs, 0.018 (±0.047) for 23 z>1 SFGs with $Z_O$>8.76, 0.2 (±0.047) for 14 z>1 SFGs with 8.76>$Z_O$>8.62, and 0.34 (±0.045) for 7 z>1 SFGs with $Z_O$<8.62. The numbers in parentheses are the 1σ uncertainties of the mean. The black diagonal black arrows in the upper right show in which direction the data points would move if the true conversion factor is half or twice the Galactic value.

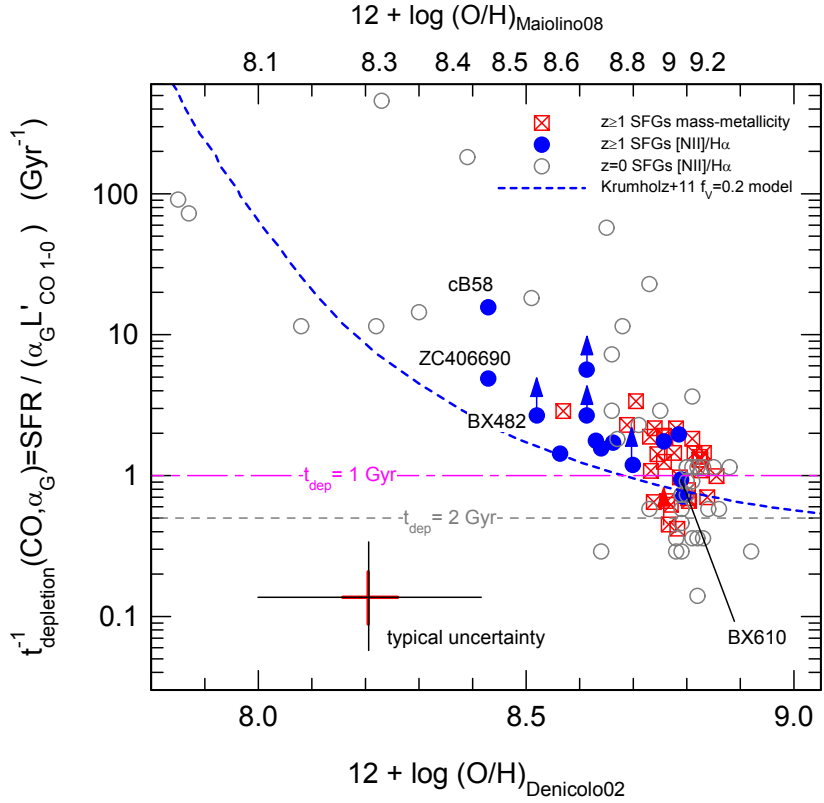

Figure 4. Dependence of the molecular gas depletion rate for a Galactic conversion factor (including helium), $(t_{depletion}(CO, \alpha_G))^{-1} = SFR/(\alpha_G L'_{CO\,1-0})$, on gas phase oxygen abundance on the Denicolo et al. (2002) scale (bottom) and the Maiolino et al. (2008) scale (top). The molecular gas mass is inferred from the observed CO 3-2/2-1 flux/luminosity from equations (1) and (2) with $R_{13}=2$ (1.5 for the eyelash) or $R_{12}=1.16$, and $\alpha_{CO\,1-0}=\alpha_G=4.36$. The blue filled circles denote CO detections or $3\sigma$ upper limits of z=1-2.5 SFGs with individual determinations of the oxygen abundance based on the [NII]/H$\alpha$ ratio, the Pettini & Pagel (2004) relation ($Z_O= 12 + \log(O/H)=8.9 +0.57 \log(F([NII]/F(H\alpha)))$, and then converted to the Denicolo et al. (2002) calibration scale by using the conversion functions given in Kewley & Ellison (2008). The crossed red squares mark CO detections/upper limits with metallicities inferred from the stellar mass-metallicity relation of Erb et al. 2006a, Liu et al. 2008 and Buschkamp et al. 2011 in prep.), again converted to the Denicolo et al. scale. Open grey circles denote the compilation of z~0 SFGs from Krumholz, Leroy & McKee (2011), for which metallicities were derived from [NII]/Hα ratios in the literature, and then again converted to the Denicolo et al. (2002) calibration. The best fit depletion time scales for near solar metallicity SFGs are $t_{depletion}$~2 Gyr at z~0 (Bigiel et al. 2008, 2011, Leroy et al. 2008) and ~1 Gyr at z~1-2.5 (Tacconi et al. 2010, this paper), which are shown as dashed horizontal lines. The large cross at the bottom denotes the typical statistical (red) and systematic (black) rms errors. The blue dashed line is a theoretical prediction of the dependence of $(t_{depletion})^{-1}$ on $Z_O$ for a molecular cloud of constant column density similar to that in Milky Way GMCs (~80 $M_\odot$ pc$^{-2}$, N(H)=7.5x10$^{21}$ cm$^{-2}$) and a ratio of UV radiation field to density similar to that in the solar neighborhood, for a star formation cloud filling factor $f_V$=0.2 (Krumholz et al. 2011).

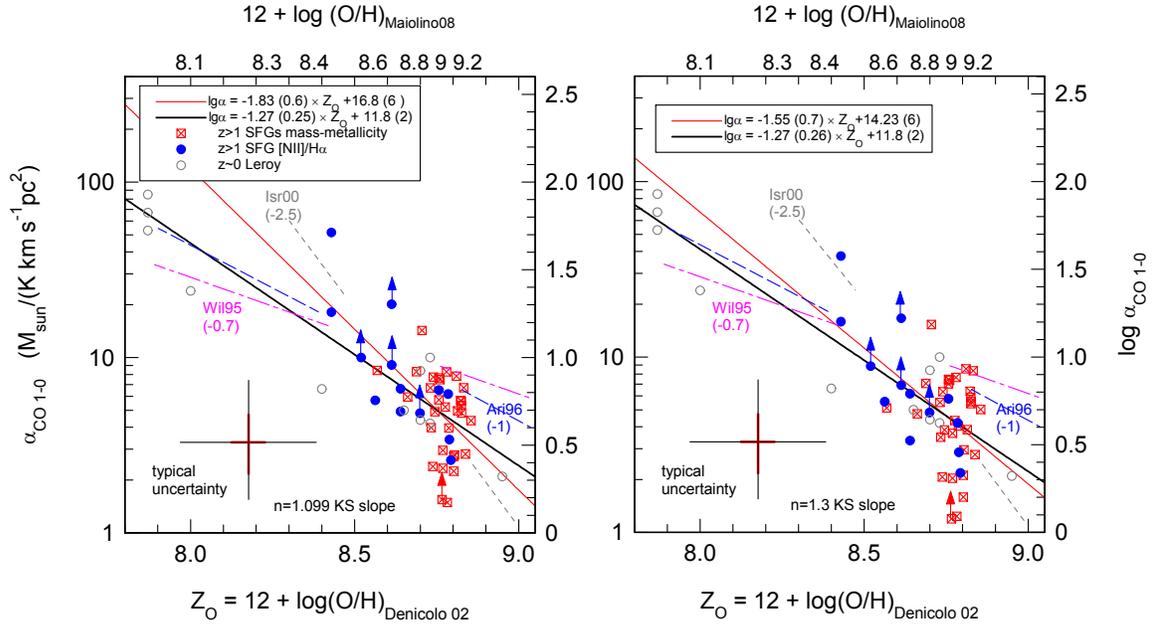

Figure 5. Inferred dependence of the CO 1-0 luminosity to molecular gas mass, conversion factor ($\alpha_{CO\,1-0}$) on gas phase oxygen abundance. The molecular gas mass (including helium) for the high-z galaxies is computed from the best-fit $z \geq 1$ KS-relation. The left inset is for our favored KS-slope of n=1.1 in Figure 3 and equation (2). The right inset is for n=1.3. The symbols for the high-z data points are the same as in Figure 4. Grey circles denote the determinations of $\alpha_{CO\,1-0}$ from joint fits to spatially resolved HI, CO and dust data in z~0 SFGs by Leroy et al. (2011). The large black (red) cross in the lower left denotes the typical systematic (statistical) rms errors. The trends of the Leroy et al. sample are in broad agreement with earlier work of Wilson (1995, magenta long-short dashes), Arimoto et al. (1996, long blue dashes) and Israel (2000, grey dots). The best linear fit in the log($\alpha_{CO\,1-0}$)-log(metallicity) plane to the all z~0 and z>1 SFGs is given by the continuous black line: log $\alpha_{CO\,1-0}$= 12 (±2) -1.3 (±0.26) $Z_O$ (for both choices of n), where the quoted errors are 2σ statistical fit uncertainties. The best fit to only z≥1 SFGs is given by the continuous red line: log $\alpha_{CO\,1-0}$= 17 (± 6) -1.8 (±0.6) $Z_O$ for n=1,099 and log $\alpha_{CO\,1-0}$= 14 (± 6) -1.6 (±0.6) $Z_O$ for n=1.3.

**Table 1. Star formation rates and depletion time scales: comparison of observed values compared to predictions of FMR and KS relations (for given metallicity and stellar mass)**

|  | ZC406690 | cB58 | BX482 | BX389 |
|---|---|---|---|---|
| z | 2.2 | 2.73 | 2.26 | 2.17 |
| $12+\log(O/H)_{D02}$ | 8.43 | 8.43 | 8.52 | 8.61 |
| $M_*$ ($10^{10}$ $M_\odot$) | 4.3 | 0.5 | 1.8 | 6.9 |
| $R_{1/2}$ (kpc) | 6.3 | 1.5 | 4.2 | 4.2 |
| SFR ($M_\odot$ yr$^{-1}$) (from H$\alpha$ + Calzetti) | 290 | 63 | 120 | 197 |
| $(t_{depletion})^{-1}$ ($\alpha$=4.36) (Gyr$^{-1}$) | 4.9 (±1.8) | 15.7 (±5) | >2.7 (3$\sigma$) | >5.7 (3$\sigma$) |
| from FMR/KS: $\beta$=0.31, n=1.1 |  |  |  |  |
| SFR | 1.6e4 | 19 | 100 | 1.1e3 |
| $(t_{depletion})^{-1}$ (Gyr)$^{-1}$ | 1.7 | 1.2 | 1.2 | 1.5 |
| from FMR/KS: $\beta$=0.37, n=1.1 |  |  |  |  |
| SFR | 6.9e3 | 64 | 65 | 910 |
| $(t_{depletion})^{-1}$ (Gyr)$^{-1}$ | 2.6 | 1.4 | 1.2 | 1.5 |
| from FMR/KS: $\beta$=0.25, n=1.1 |  |  |  |  |
| SFR | 5.4e4 | 8.3 | 200 | 1.3e3 |
| $(t_{depletion})^{-1}$ (Gyr)$^{-1}$ | 1.9 | 1.2 | 1.3 | 1.5 |
| from FMR/KS: $\beta$=0.31, n=1.3 |  |  |  |  |
| SFR | 1.6e4 | 19 | 100 | 1.1e3 |
| $(t_{depletion})^{-1}$ (Gyr)$^{-1}$ | 4.1 | 1.7 | 1.5 | 2.6 |

4545